\documentclass[reprint, amsmath,amssymb, aps,pra,]{revtex4-2}
\usepackage{graphicx}
\usepackage{dcolumn}
\usepackage{bm}
\usepackage{mathtools}
\usepackage[utf8]{inputenc}
\usepackage[T1]{fontenc}
\usepackage{booktabs, array, float, tabularx, lipsum, amsmath,multirow}
\usepackage{siunitx, xcolor}
\usepackage[version=4]{mhchem}
\usepackage {nopageno} 
\graphicspath{{figs/}{figsgaoerb/}} 
\usepackage[colorlinks,linkcolor=blue,anchorcolor=blue,citecolor=blue]{hyperref}
\usepackage{mathptmx}
\DeclareSymbolFont{cmletters}{OML}{cmm}{m}{it}
\DeclareMathSymbol{\Gamma}{\mathalpha}{cmletters}{0}
\DeclareMathSymbol{\Delta}{\mathalpha}{cmletters}{1}
\DeclareMathSymbol{\Theta}{\mathalpha}{cmletters}{2}
\DeclareMathSymbol{\Lambda}{\mathalpha}{cmletters}{3}
\DeclareMathSymbol{\Xi}{\mathalpha}{cmletters}{4}
\DeclareMathSymbol{\Sigma}{\mathalpha}{cmletters}{6}
\DeclareMathSymbol{\Upsilon}{\mathalpha}{cmletters}{7}
\DeclareMathSymbol{\Phi}{\mathalpha}{cmletters}{8}
\DeclareMathSymbol{\Psi}{\mathalpha}{cmletters}{9}
\DeclareMathSymbol{\Omega}{\mathalpha}{cmletters}{10}
\DeclareMathSymbol{\alpha}{\mathord}{cmletters}{11}
\DeclareMathSymbol{\beta}{\mathord}{cmletters}{12}
\DeclareMathSymbol{\gamma}{\mathord}{cmletters}{13}
\DeclareMathSymbol{\delta}{\mathord}{cmletters}{14}
\DeclareMathSymbol{\epsilon}{\mathord}{cmletters}{15}
\DeclareMathSymbol{\zeta}{\mathord}{cmletters}{16}
\DeclareMathSymbol{\eta}{\mathord}{cmletters}{17}
\DeclareMathSymbol{\theta}{\mathord}{cmletters}{18}
\DeclareMathSymbol{\iota}{\mathord}{cmletters}{19}
\DeclareMathSymbol{\kappa}{\mathord}{cmletters}{20}
\DeclareMathSymbol{\lambda}{\mathord}{letters}{21}
\DeclareMathSymbol{\mu}{\mathord}{cmletters}{22}
\DeclareMathSymbol{\nu}{\mathord}{cmletters}{23}
\DeclareMathSymbol{\xi}{\mathord}{cmletters}{24}
\DeclareMathSymbol{\pi}{\mathord}{cmletters}{25}
\DeclareMathSymbol{\rho}{\mathord}{cmletters}{26}
\DeclareMathSymbol{\sigma}{\mathord}{cmletters}{27}
\DeclareMathSymbol{\tau}{\mathord}{cmletters}{28}
\DeclareMathSymbol{\upsilon}{\mathord}{cmletters}{29}
\DeclareMathSymbol{\phi}{\mathord}{cmletters}{30}
\DeclareMathSymbol{\chi}{\mathord}{cmletters}{31}
\DeclareMathSymbol{\psi}{\mathord}{cmletters}{32}
\DeclareMathSymbol{\omega}{\mathord}{cmletters}{33}
\DeclareMathSymbol{\varepsilon}{\mathord}{cmletters}{34}
\DeclareMathSymbol{\vartheta}{\mathord}{cmletters}{35}
\DeclareMathSymbol{\varpi}{\mathord}{cmletters}{36}
\DeclareMathSymbol{\varrho}{\mathord}{cmletters}{37}
\DeclareMathSymbol{\varsigma}{\mathord}{cmletters}{38}
\DeclareMathSymbol{\varphi}{\mathord}{cmletters}{39}
\begin{document}
\renewcommand\arraystretch{1}
\preprint{APS/123-QED}
\title{Experimental evidence on the restricted contextual advantage in mirror-symmetric state discrimination}

\author{Xuan Fan}
\author{Ya Xiao}%
\email{xiaoya@ouc.edu.cn}
\author{Yongjian Gu}
\email{yjgu@ouc.edu.cn}
\affiliation{%
 College of Physics and Optoelectronic Engineering, Ocean University of China, Qingdao City, Shandong Province, People's Republic of China, 266100\\
}%
\date{\today}

\begin{abstract}
Contextuality is well known as a vital resource for locating the boundary between classical and quantum theories, as well as identifying tasks showing quantum advantage. In a surge of recent works [Schmid and Spekkens, Phys.Rev.X 8, 011015 (2018); Mukherjee, Naonit and Pan, Phys.Rev.A 106, 012216 (2022); Flatt, Lee, Carceller, Brask and Bae, PRX QUANTUM 3, 030337 (2022)], it has also been shown that contextuality is the crucial resource in quantum state discrimination (QSD) tasks, including minimum error discrimination (MED) and maximum confidence discrimination (MCD), together with many other figure-of-merits. Despite the fundamental progress made by those aforementioned works, none of them mention about how to realize their fancy proposals, which is doubtlessly necessary for the final goal of applying this resource in real QSD tasks. 
In this paper, we report the first experimental demonstration of contextual advantage in both MED and MCD for three mirror-symmetric states using interferometric quantum walk, which can be easily generalized to any figure-of-merit in QSD. Our experiment agrees well with the result of theoretical simulation, and also shows the great potentiality of leveraging this method to explore a simpler version for the witness of contextuality, as well as demonstrating quanutm advantage of various tasks that require QSD. 

\end{abstract}
\maketitle
\section{\label{Intro}Introduction}

Contextuality, as a basic non-classicality of foundational importance, not only plays a crucial role in helping to understand the difference between quantum and classical world, but also serves as an important resource in quantum information science\cite{budroni2022kochen}. In the past years, researches on the applications of contextuality have garnered unprecedented attention, showcasing its various intriguing characteristics and potential practical significance in quantum computation and cryptography\cite{spekkens2009preparation,hoban2011non,raussendorf2013contextuality,howard2014contextuality}. Up to now, researches have proven that quantum advantage can be shown by the failure of noncontextuality in a broad range of applications, such as fault-tolerant quantum computation\cite{marques2014experimental,delfosse2015wigner,mansfield2018quantum}, quantum random access codes\cite{chailloux2016optimal,ambainis2019parity,gatti2023random}, magic state distillation\cite{bravyi2005universal,dawkins2015qutrit,prakash2020contextual}, etc. In recent years, researchers have found that contextuality is also the key resource for a basic quantum information processing task known as state discrimination\cite{schmid2018contextual,flatt2022contextual,mukherjee2022discriminating}. \par
Quantum state discrimination (QSD) plays a vital role in quantum information theory and quantum communication tasks, mainly focused on the effective determination of quantum states according to a set of measured probabilities with known prior distributions. However, as the superposition principle indicates, there must be states being non-orthogonal and thus having a non-zero overlap, which makes them impossible to be perfectly distinguished with certainty, leading to no-cloning theorem\cite{wootters1982single}. Therefore it is necessary to study upon the optimization of non-orthogonal state discrimination based on various strategies, such as unambiguous discrimination (UD)\cite{ivanovic1987differentiate,dieks1988overlap,peres1988differentiate}, minimum error discrimination (MED)\cite{barnett2001minimum,andersson2002minimum}, and maximum confidence discrimination (MCD)\cite{croke2006maximum,mosley2006experimental}.\par
The fact that contextuality plays an important role in quantum state discrimination has stimulated broad interests in further exploring this particular field of research. In the work aforementioned \cite{schmid2018contextual}, both contextual and non-contextual theories have been characterised  
in two-state MED. But it has a presumption that two states should have equal prior probabilities, which is later generalized to two-state with different probabilities, and to three-state discrimination\cite{mukherjee2022discriminating}, as well as to other QSD scenarios such as MCD, and UD\cite{flatt2022contextual}. However, up to now few work has been focused upon the realization of those tasks, which is doubtlessly essential for their final application. In this work, we put forward with an optical setup for MED and MCD scenarios, which can be even further generalized to realize any measurement strategy. The method of utilizing quantum walk as a generalized measuring device has first been proposed in \cite{kurzynski2013quantum}, and soon gone through experimental verification as well as wide application\cite{bian2015realization,zhao2015experimental,wang2023generalized}. 
Using interferometric quantum walk, we demonstrate the contextual advantage in both two scenarios aforementioned of three mirror-symmetric states for the first time. In both scenarios, an obvious violation of the non-contextual bound can be observed, with the results fitting in good agreement with quantum predictions. Our work provides a practical and generalizable pathway for realizing contextual advantage given any figure-of-merit, and also shows the potentiality of utilizing this method to further explore the performance of contextuality in QSD scenarios based on optical system.
\section{\label{Theory}Theory}
\subsection{\label{MEDMCD}Two figure-of-merits for QSD: MED and MCD}
The fundamental goal of QSD is to design a measurement that optimally discriminates the given states by a set of positive operator valued measures (POVMs). In other words, it is possible to find a set of POVM operators describing the correspondent measurement, given an optimal figure-of-merit. 
So far, several figure-of-merits have been developed for the optimization of QSD, including UD, which identifies a state unambiguously at the cost of some inconclusive results, but suits only independent states without extra copies\cite{chefles2001unambiguous}; MED, which minimizes the probability of making a guessing error in state identification; and MCD, which is introduced as a more general analogy to UD, but suits also those states that are linearly dependent. In this paper we will only talk about MED and MCD, since UD is only possible when the states to be distinguished are linearly independent to each other. For the convenience of demonstration, without loss of generality we take here mirror-symmetric state discrimination for both MED and MCD tasks as example. \par
\begin{figure}[t]
\begin{center}
\includegraphics[width=0.6\columnwidth]{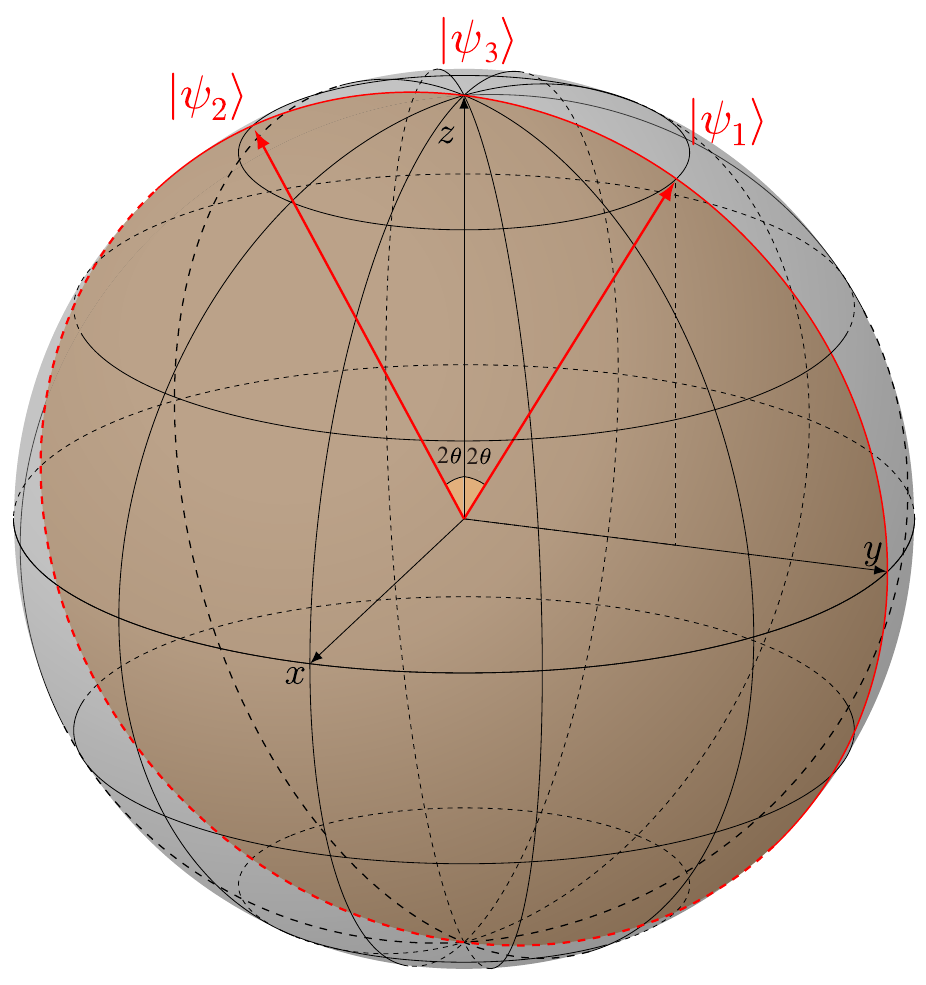}
\caption{Representation of three mirror-symmetric states on the $x-z$ plane of Bloch sphere. The red circular circle on the marigold surface denotes feasible range of states, where $|\psi_3\rangle$ acts as the central axis for mirror symmetry between $|\psi_1\rangle$ and$|\psi_2\rangle$, both of which are 2$\theta$ to $|\psi_3\rangle$. In our real experiment, different initial coin states are prepared by correspondently taking $\theta\in(0,\frac{\pi}{2})$.}
\label{Bloch}
\end{center}
\end{figure}

The three mirror-symmetric states used in this paper take the following form, as shown in Fig. \ref{Bloch}. 
\begin{equation}
    \begin{aligned}
&|\mathcal{\psi}_1\rangle=\cos\theta|0\rangle+\sin\theta|1\rangle,\\
        &|\psi_2\rangle=\cos\theta|0\rangle-\sin\theta|1\rangle,\\
        &|\psi_3\rangle=|0\rangle,
    \end{aligned}\label{State}
\end{equation}
and those three states are taken with prior probabilities $p_1=p_2=p$ and $p_3=1-2p$, where $0<p<1/2$ and $0<\theta<\pi/2$. \par
For mirror-symmetric three-state MED, the optimal POVM choice is determined as follows by the correspondent figure-of-merit. Supposedly $p<\frac{1}{2+\cos\theta(\cos\theta+\sin\theta)}$, the full three-element POVM would be available: 
\vspace{0.1cm}
\begin{equation}\label{MED3}
\textstyle\Pi_1 =\scriptstyle\left(\begin{array}{cc}
 \frac{\mu^2}{2} &  \frac{\mu}{2} \\
 \frac{\mu}{2} &  \frac{1}{2}
\end{array}\right),\;
\textstyle\Pi_2 =\scriptstyle \left(\begin{array}{cc}
 \frac{\mu^2}{2} &  \scriptstyle{-}\textstyle\frac{\mu}{2} \\
 \scriptstyle{-}\textstyle\frac{\mu}{2} &  \frac{1}{2}
\end{array}\right),\;
\displaystyle\Pi_3 = \scriptstyle\left(\begin{array}{cc}
\scriptstyle 1-\mu^2 &\scriptstyle 0 \\
\scriptstyle0 & \scriptstyle0
\end{array}\right), 
\end{equation}
\vspace{0.1cm}
where $\mu=\frac{p\cos\theta\sin\theta}{1-p(2+\cos^2\theta)}$, and we have $\Pi_1+\Pi_2+\Pi_3=\mathbb{I}$. \par
However, noticeably POVM element $\Pi_3$ will become unphysical when $1-\mu^2>0$. Thus it was suggested in \cite{andersson2002minimum} that if  $p\ge\frac{1}{2+\cos\theta(\cos\theta+\sin\theta)}$, the best solution degrades into an optimal two-state discrimination between $|\psi_1\rangle$ and $|\psi_2\rangle$, where the three-element POVM becomes a projetive measurement:
\begin{equation}\label{MED2}
    \begin{aligned}
        &\Pi_1 = \left(\begin{array}{cc}
\frac{1}{2} &  \frac{1}{2} \\
 \frac{1}{2} &  \frac{1}{2}
\end{array}\right), \;\Pi_2 = \left(\begin{array}{cc}
 \frac{1}{2} &  -\frac{1}{2} \\
 -\frac{1}{2} &  \frac{1}{2}
\end{array}\right).
    \end{aligned}
\end{equation}\par
\vspace{0.1cm}
Combining Eqs. (\ref{MED3})-(\ref{MED2}), we have the maximal success probability for MED strategy: 
\vspace{-0.2cm}
\begin{equation}
    \begin{aligned}
        S_Q &= \sum_0^ip_i\rm Tr( \rho_\textit{i} \it E_i\rm )\\
        &=  \begin{cases}
        p(1+\sin2\theta), &  \scriptstyle p\geq \frac{1}{2+\cos\theta(\cos\theta+\sin\theta)},\\
         \frac{(1-2p)(p\sin^2\theta+1-2p-p\cos^2\theta)}{2+\cos\theta(\cos\theta+\sin\theta)}, & \scriptstyle p< \frac{1}{2+\cos\theta(\cos\theta+\sin\theta)}.\\
    \end{cases}
    \end{aligned}
\end{equation}
\par
The same logic also applies for MCD strategy, where a four-element POVM would be needed for a universal solution: 
        \begin{center}
        \vspace{-0.6cm}
\begin{equation}\label{MCD4}
    \begin{aligned}
\Pi_1 &= \left(\begin{array}{cc}
 \frac{\nu^2\xi}{2} &  \frac{\nu\xi}{2} \\
 \frac{\nu\xi}{2} &  \frac{\xi}{2}
\end{array}\right),
        \Pi_2 = \left(\begin{array}{cc}
\frac{\nu^2\xi}{2} &  -\frac{\nu\xi}{2} \\
 -\frac{\nu\xi}{2} &  \frac{\xi}{2}
\end{array}\right),\; \\
&\Pi_3 = \left(\begin{array}{cc}
\scriptstyle 1-\nu^2\xi & \scriptstyle0 \\
\scriptstyle 0 & \scriptstyle0
\end{array}\right), \;
\Pi_4 = \left(\begin{array}{cc}
 \scriptstyle0 &  \scriptstyle0 \\
 \scriptstyle0 &  \scriptstyle 1-\xi
\end{array}\right).
    \end{aligned}
\end{equation}
\end{center}\par
However, within the range of $p<\frac{1}{2\sin^2\theta+\sin2\theta}$, it is possible for us to further simplify Eqs. (\ref{MCD4}) into a three-element POVM expression as follows:
\begin{equation}\label{MCD3}
    \begin{aligned}
        &\Pi_0 = \scriptstyle\left(\begin{array}{cc}
\scriptstyle 1-\nu^2 &\scriptstyle 0 \\
\scriptstyle0 & \scriptstyle0
\end{array}\right), \;
\textstyle\Pi_1 =\scriptstyle\left(\begin{array}{cc}
 \frac{\nu^2}{2} &  \frac{\nu}{2} \\
 \frac{\nu}{2} &  \frac{1}{2}
\end{array}\right),\: \textstyle\Pi_2 =\scriptstyle \left(\begin{array}{cc}
 \frac{\nu^2}{2} &  \scriptstyle{-}\textstyle\frac{\nu}{2} \\
\scriptstyle{-}\textstyle\frac{\nu}{2} &  \frac{1}{2}
\end{array}\right),
    \end{aligned}
\end{equation}
where $\nu=\frac{p\sin 2\theta}{1-2p\sin^2\theta}$. As can be seen, the completeness condition is also satisfied. 
Similarly, from Eqs. (\ref{MCD4})-(\ref{MCD3}), we can deduce the maximal success probability for MCD strategy: 
\begin{equation}
    \begin{aligned}
        C_Q &= \frac{p_1 Tr[\rho_1E_1]}{Tr[\rho E_1]}
        =  \frac{1+2p\cos\theta}{2-4p\sin^2\theta}.\\
    \end{aligned}
\end{equation}
\par
The success probabilities of both MED and MCD allowed in the non-contextual ontological model are bounded by:
\begin{equation}
    \begin{aligned}
        &S_\lambda = \begin{cases}
        1-(1-2p)\cos^2\theta-p\cos^22\theta, & p\geq \frac{1}{3},\\
        1-p\cos^2\theta-p\cos^22\theta, & p< \frac{1}{3}.\\
    \end{cases}\\
        &C_\lambda = \frac{1}{1+\cos^22\theta+(\frac{1}{p}-2)\cos^2\theta}.
    \end{aligned}
\end{equation}\par
Any experimental violation of the bounds above would serve as a witness for contextuality, regardless of the validity of quantum theory. On the other hand, given the maximum success probabilities under contextual and non-contextual theories, we have a clear image of where contextuality shows quantum advantage, as shown in Fig. \ref{MEDresult}.
\subsection{\label{Contextual}Construction of POVM for MED and MCD through Quantum Walk}
No matter which specific figure-of-merit a QSD task will take, obviously the first step for any of them is to realize the POVM operators that they would require. Here we realize the required POVM through a method called one-dimensional discrete-time quantum walk\cite{kurzynski2013quantum,zhao2015experimental,bian2015realization}. \par
One-dimensional discrete-time quantum walk, hereinafter referred to briefly as quantum walk for simplicity, describes the coin-dependent evolution of a quantum particle on a one-dimensional lattice. 
There are two degrees of freedom in a typical quantum walk model: a walker, namely the position of the quantum particle $x \in \{...,-1,0,1,...\}$ , and a coin $|c\rangle=\{\alpha,\beta\}^T$ carried by the walker, which gets flipped before taking each step. In the $n$th step,  the site-dependent coin rotation $C_{n}^{(x)}$ is applied to the coin when the walker is at position $x$, which is followed by a conditional shift $T=\sum_x|x+1\rangle\langle x|\otimes|0\rangle\langle 0|+\sum_x|x-1\rangle\langle x|\otimes|1\rangle\langle 1|$ due to the previous coin rotation. The unitary operation for the $n$th step can be expressed as $U_n=T\sum_x|x\rangle\langle x|\otimes C_{n}^{(x)}$.
POVM is realized through the combination of projective walker measurement and coin-walker entanglement, while the indirect control of $|c\rangle$ on particle evolution yields the inference from $|x\rangle$ to $|c\rangle$. \par
A more detailed procedure of quantum walk is as follows.\par
1. Initiate the quantum walk at position $x = 0$ with the coin state corresponding to the qubit state one wants to measure. \par
2. Set $ i\coloneqq1$.  \par
3. While $i<n$ do the following: \par
(1) Apply coin operation $C^{(1)}_i$ at position $x = 0$ and identity matrix elsewhere and then apply translation operator $ T$ .\par
(2) Apply coin operation $C^{(2)}_i$  at position $x = 1$, apply $\textrm{NOT} = \left(\begin{array}{cc}
0 &  1 \\
1 &  0
\end{array}\right)$ at position $x =-1$, and identity matrix $\text{I} = \left(\begin{array}{cc}
1 &  0 \\
0 &  1
\end{array}\right)$ elsewhere, then apply translation operator $T$ . \par
(3) $i\coloneqq i + 1$.\par
Here we follow the general algorithm as proposed in \cite{kurzynski2013quantum}. By repeating the procedures shown above, we can realize any set of rank 1 or rank 2 POVM elements. For mirror-symmetric states as shown in Eq. (\ref{State}), the site-dependent coin rotations that satisfy MED strategy Eq. (\ref{MED3}) and Eq. (\ref{MED2}) are as follows:
\begin{equation}
\begin{aligned}\label{MEDD}
&\mathrm{If}\hspace{0.1cm}p<\frac{1}{2+\cos\theta(\cos\theta+\sin\theta)},\hspace{0.1cm}C_1^{(1)}=\mathbb{I}, \\
&C_1^{(2)}=\left(\begin{array}{cc}
\scriptstyle\sqrt{1-(\frac{p\cos\theta\sin\theta}{1-p(2+p\cos^2\theta)})^2} &  \frac{p\cos\theta\sin\theta}{1-p(2+p\cos^2\theta)} \\
\frac{p\cos\theta\sin\theta}{1-p(2+p\cos^2\theta)} &  \scriptstyle-\sqrt{1-(\frac{p\cos\theta\sin\theta}{1-p(2+p\cos^2\theta)})^2}
\end{array}\right), \\
&C_2^{(1)}=\sqrt{\frac{1}{2}}\left(\begin{array}{cc}
1 &  1 \\
1 &  -1
\end{array}\right),\hspace{0.1cm}C_2^{(2)}=\mathbb{I};\\
&\mathrm{Else\hspace{0.1cm}if}\hspace{0.1cm}p\ge\frac{1}{2+\cos\theta(\cos\theta+\sin\theta)},\hspace{0.1cm}C_1^{(1)}=\mathbb{I},\\
&C_1^{(2)}=\left(\begin{array}{cc}
0 &  1 \\
1 &  0
\end{array}\right), \hspace{0.1cm}C_2^{(1)}=\sqrt{\frac{1}{2}}\left(\begin{array}{cc}
1 &  1 \\
1 &  -1
\end{array}\right),\hspace{0.1cm}C_2^{(2)}=\mathbb{I}.    
\end{aligned}
\end{equation}\par
While similarly for MCD strategy, the site-dependent coin rotations satisfy Eq. (\ref{MCD3}) should be like: 
\begin{equation}
\begin{aligned}\label{MCDD}
&C_1^{(1)}=\mathbb{I}, \\
&C_1^{(2)}=\left(\begin{array}{cc}
\scriptstyle\sqrt{1-(\frac{p\sin 2\theta}{1-2p\sin^2\theta})^2} &  \frac{p\sin 2\theta}{1-2p\sin^2\theta} \\
 \frac{p\sin 2\theta}{1-2p\sin^2\theta} &  \scriptstyle-\sqrt{1-(\frac{p\sin 2\theta}{1-2p\sin^2\theta})^2}
\end{array}\right), \\
&C_2^{(1)}=\sqrt{\frac{1}{2}}\left(\begin{array}{cc}
1 &  1 \\
1 &  -1
\end{array}\right),\hspace{0.1cm}C_2^{(2)}=\mathbb{I}.    
\end{aligned}
\end{equation}

Now our goal is to experimentally achieve the single-qubit POVM for both MED and MCD of three mirror-symmetric states by implementing a well-designed quantum walk, so as to discriminating the initial coin states with either minimized error or maximized confidence. It is worth noticing that with a quantum walk properly engineered for a given figure-of-merit, the POVM of any QSD can be realized using the same logic.
\section{\label{Experiment}Experiment}
Experimentally, we choose to encode the coin with photonic polarization, and the walker with photonic path. 
\begin{figure}[t]
\begin{center}
\includegraphics[width=1\columnwidth]{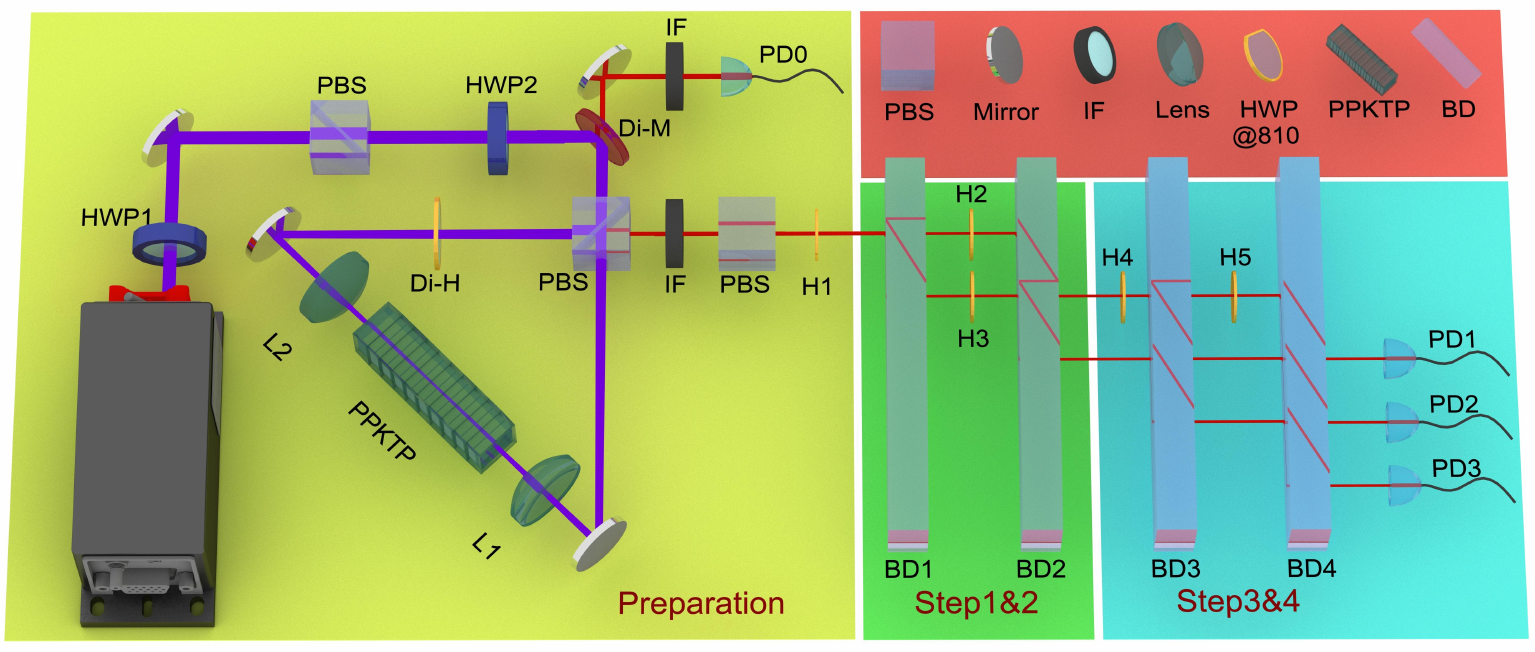}
\caption{Schematic of the experimental setup. Initial coin states as shown in Eq.\hspace{0.05cm}(\ref{State}) are prepared in the yellow part, where heralded single photons are generated via spontaneous parametric down-conversion (SPDC) and steered into 3 different states by the combination of a polarizing beam splitter (PBS) and a half-wave plate (HWP). After preparation, the single photons undergo two steps of quantum walk, indicated by the green and blue parts respectively. Each step includes coin operations realized by HWPs and conditional shift operations realized by beam displacers (BDs). Finally, heralded photons are detected by avalanche photodiodes (APDs), each of which corresponds to an outcome of a POVM element, via a coincidence with the triggered photons at PD0. }
\label{setup}
\end{center}
\end{figure}
Down-converted polarization-separable photons are generated via the spontaneous parametric down conversion (SPDC) process by pumping the type-II cut PPKTP crystal in a Sagnac interferometer with 405nm ultraviolet laser. Due to the imperfect efficiency of  SPDC, we use two interference filters of 3nm bandwidth to filter photon pairs, after which one side is sent to avalanche photodiode (APD) PD0, which heralds the presence of a single photon state  $|H\rangle$ on the other side; while the other side, at single photon state $|H\rangle$, is sent to perform quantum walking. In quantum walk, the initial coin states $|\psi\rangle=\cos\theta|H\rangle+\sin\theta|V\rangle$ are prepared by passing the single photons through a polarizing beam splitter (PBS), a half-wave plate (HWP) in the correspondent configuration. Note that here we only talk about the special case of mirror-symmetric state discrimination, for other states an extra quater wave plate (QWP) might be needed. PD1, PD2 and PD3 are three APDs, each of which has a response in accordance with the measurement outcome of the correspondent POVM element.


\subsection{\label{MEDexp}POVM for three mirror-symmetric state MED}
\begin{figure}[t]
\begin{center}
\includegraphics[width=1\columnwidth]{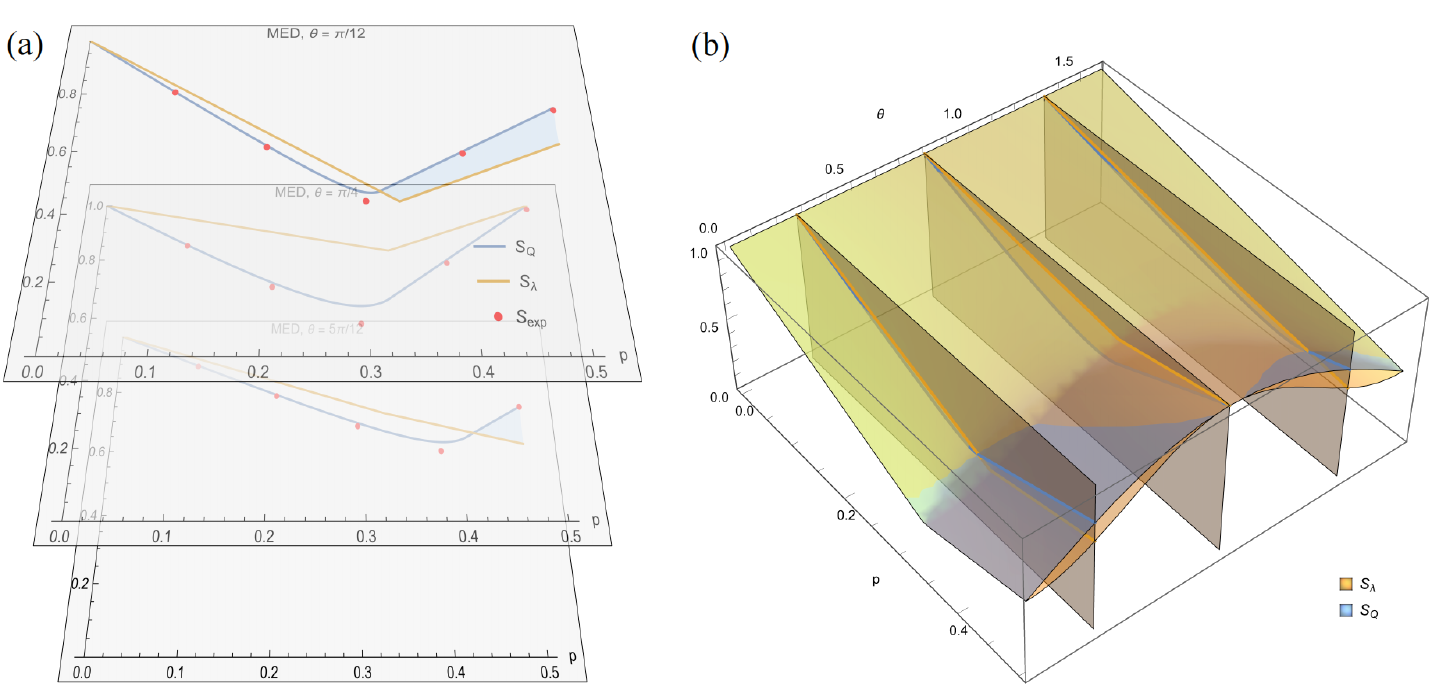}
\caption{Results for MED of three mirror-symmetric states averaged over 30 experimental runs. (a) Comparison between experimental result and theoretical predictions for  $p\in\{0.1, 0.2, 0.3, 0.4, 0.5\}$ and $\theta\in\{\frac{\pi}{12},\frac{\pi}{4},\frac{5\pi}{12}\}$. Blue and yellow curves represent quantum and non-contextual bounds, while the region where contextual advantage can be seen is filled with light blue. Red dots are results over 30 experimental runs, the error bars of which are not visible on this scale. As can be seen, our result has a good fit with the quantum prediction, while violating the classical bound of non-contextual hidden variable theory in certain regions. 
(b) An overall picture of quantum and non-contextual bounds over $p\in(0,0.5]$ and $\theta\in(0,\frac{\pi}{2})$. The advantage of contextuality can be witnessed in the blue region, which is dependent on both $p$ and $\theta$. For the region of $p>0.25$, there are two parts that demonstrate contextuality, with $\theta=\frac{\pi}{4}$ as an exceptional point.}
\label{MEDresult}
\end{center}
\end{figure}
With the experimental setup in Fig. \ref{setup}, we are able to construct the POVM for three mirror-symmetric state MED. For a more inclusive understanding of the quantum superior region, where contextuality shows advantage against classical predictions of non-contextual hidden variable theory, we take a series of $\theta\in\{\frac{\pi}{12},\frac{\pi}{4},\frac{5\pi}{12}\}$ and $p\in\{0.1, 0.2, 0.3, 0.4, 0.5\}$ into our investigation. Now let's take a special case with $\{p,\theta\}=\{0.3, \frac{\pi}{12}\}$ as an example to show the specific settings of our setup. Substitute the value into Eq. (\ref{State}), we have:
\begin{equation}
    \begin{aligned}
&|\mathcal{\psi}_1\rangle=\frac{1+\sqrt{3}}{2\sqrt{2}}|0\rangle+\frac{\sqrt{3}-1}{2\sqrt{2}}|1\rangle,\\
        &|\psi_2\rangle=\frac{1+\sqrt{3}}{2\sqrt{2}}|0\rangle+\frac{1-\sqrt{3}}{2\sqrt{2}}|1\rangle,\\
        &|\psi_3\rangle=|0\rangle,
    \end{aligned}\label{State1}
\end{equation}
which can be prepared with the rotation of H1 by $\theta_{H_1}=\{\frac{\pi}{24}, \scriptstyle-\textstyle\frac{\pi}{24}, 0\}$. 
The coin operators $C_1^{(1)}$ and $C_2^{(2)}$ as shown in Eq. (\ref{MEDD}) are identity matrices which correspond to do-nothing operation, while $C_1^{(2)}$ and $C_2^{(1)}$ are realized by respectively rotating H3 and H4 to $\theta_{H_{3}}\approx0.038\pi$ and $\theta_{H_{4}}=\frac{\pi}{8}$. Both H2 and H5 are rotated to $\theta_{H_{2,5}}=\frac{\pi}{4}$ to perform NOT operation. By repeating the above-mentioned procedures, we can construct the POVM needed for other $\{p,\theta\}$ sets, and reach to Fig. \ref{MEDresult} after scanning all $p$ and $\theta$ needed. \par
Our experimental results are shown as red points in Fig. \ref{MEDresult}(a), where the contextual and non-contextual bounds are drawn in blue and yellow, respectively. The light blue region denotes where contextual prediction surpasses its non-contextual counterpart, i.e. the region in which contextual advantage can be witnessed. It is evident that our result fits well with theoretical predictions, demonstrating contextual advantage in both $\theta=\frac{\pi}{12}$ and $\theta=\frac{5\pi}{12}$. However, it's worth noting that given a large $p$, the size and depth of contextual advantageous region shows a high dependence on $\theta$, thus worth further detailed discussion. 
As Fig. \ref{MEDresult}(b) depicts, when $\theta$ increases, the contextual advantageous region grows, recedes, and totally disappears at $\theta=\frac{\pi}{4}$, after which it gets partially but not fully recovered, till its final death, showing overall a restricted contextual advantage. Noticeably, the region with a smaller $\theta$ (where $\theta<\frac{\pi}{2}$) shows greater violation of non-contextual bound (and thus the advantage of contextuality) both in its size and its depth.
\subsection{\label{MCDexp}POVM for three mirror-symmetric state MCD}
\begin{figure}[t]
\begin{center}
\includegraphics[width=1\columnwidth]{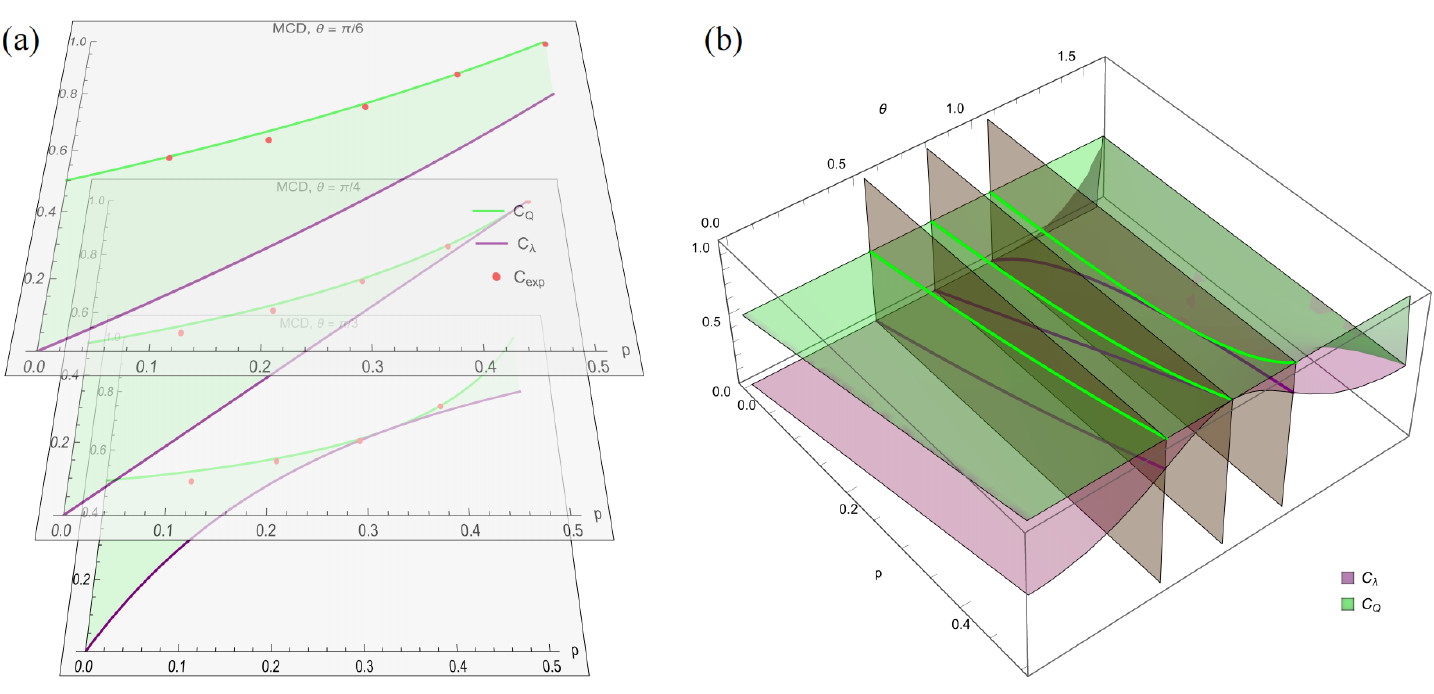}
\caption{Results for MCD of three mirror-symmetric states averaged over 30 experimental runs. (a) Comparison between experimental result and theoretical predictions for $p\in\{0.1, 0.2, 0.3, 0.4, 0.5\}$ and $\theta\in\{\frac{\pi}{6},\frac{\pi}{4},\frac{\pi}{3}\}$.
Green and purple curves represent quantum and non-contextual bounds, while the region where contextual advantage can be seen is in light green. Red dots stand for the results over 30 experimental runs, with their error bars also not visible on this scale. As can be seen, our result has a good fit with the quantum prediction, with a good violation of the classical bound of non-contextual hidden variable theory. (b) An overall picture of contextual and non-contextual bounds over $\{p,\theta\}$. Contextual advantage can be established almost across all $p$ and $\theta$ except for when $p=\frac{\csc^2\theta-2\cos2\theta}{2(3+4\cos4\theta)}$.}
\label{MCDresult}
\end{center}
\end{figure}

Similarly for MCD, we can use the same setup to construct the needed three mirror-symmetric states and correspondent optimal POVM according to the new figure of merit as shown in the previous section. Here we take as an example the special case of $\{p,\theta\}=\{0.1,\frac{\pi}{3}\}$ to demonstrate the procedures. By substituting $\theta=\frac{\pi}{3}$ into Eq.(\ref{State}), it becomes:
\begin{equation}
    \begin{aligned}
&|\mathcal{\psi}_1\rangle=\frac{1}{2}|0\rangle+\frac{\sqrt{3}}{2}|1\rangle,\\
        &|\psi_2\rangle=\frac{1}{2}|0\rangle-\frac{\sqrt{3}}{2}|1\rangle,\\
        &|\psi_3\rangle=|0\rangle,
    \end{aligned}\label{State2}
\end{equation}
the preparation of which can be realized with the rotation of H1 by $\theta_{H_1}=\{\frac{\pi}{6}, \scriptstyle-\textstyle\frac{\pi}{6}, 0\}$. The other waveplates H2-H5 are oriented at $\theta_{H_2}=\frac{\pi}{4}, \theta_{H_3}=0.016\pi, \theta_{H_4}=\frac{\pi}{8}, $ and $\theta_{H_5}=\frac{\pi}{4}$, respectively.  Fig. \ref{MCDresult} shows the result of our experiment after applying this method on the other $\{p,\theta\}$ sets. \par
As can be seen in Fig. \ref{MCDresult}(a), red points stand for the real experimental data, which is in good accordance with the theoretical predictions of quantum theory for all the $\{p,\theta\}$ tested, represented by green curves in the figure. 
Unlike its MED counterpart which possesses only a rather restricted region of contextual advantage size-wise and depth-wise, MCD strategy demonstrates much greater contextual advantage overall in both scales, showcasing its comparative superiority in tasks such as imperfect conclusive quantum teleportation\cite{neves2012quantum}. The only forbidden zone for the universal contextual advantage in MCD is when the condition of $p=\frac{\csc^2\theta-2\cos2\theta}{2(3+4\cos4\theta)}$ is satisfied.

\section{\label{summary}Conclusions}

In summary, we have experimentally realized MCD and MED strategies for mirror-symmetric states, and demonstrated the existence of contextual advantage in both cases, based on the one-dimensional discrete-time quantum walk model presented in \cite{kurzynski2013quantum}. Our work is motivated by a series of work on the performance and application of contextuality in QSD tasks \cite{schmid2018contextual,mukherjee2022discriminating,flatt2022contextual}, where the authors have theoretically discussed about the potential contextual advantage in quantum strategy in QSD scenarios of various kind, such as MED and MCD. We have put the theory forward into practice by experimentally demonstrating this advantage of contextuality in a three-state version of both strategies, which is the first of this kind according to our knowledge, making the final realization of this potential application of contextuality a step closer. \par
Our results are well fitted to the theoretical prediction of quantum theory which outperforms its classical counterpart, with a noticeable violation against the classical bound of non-contextual hidden variable theory, demonstrating $\{p,\theta\}$-dependent contextual advantage in both MCD and MED strategies for three mirror-symmetric states. According to the result, MED has a restricted advantage of contextuality, which goes after generation-expansion-deterioration-extinction-regeneration in accordance with the evolution of $\theta\in(0,\frac{\pi}{2})$. MCD, on the other hand, owns a larger non-trivial region, showing contextual advantage universally except for when $p=\frac{\csc^2\theta-2\cos2\theta}{2(3+4\cos4\theta)}$.
Furthermore, this paper unveils the potential plausibility of further applying the same method used in our work, which can be easily generalized to fit with the POVM needed in various scenarios of other QSD tasks. We believe that these kinds of experimental setups can be used in the future to unveil the advantage of contextuality as a core quantum resource in other QSD tasks, such as UD\cite{chefles1998optimum,bergou2012optimal}, mixed state discrimination\cite{rudolph2003unambiguous,zhuang2017optimum}, or discrimination with a fixed error\cite{fiuravsek2003optimal,herzog2012optimal,bagan2012optimal}.

It would also be interesting for us to further explore the following directions in the future. First, despite that our results has indeed verified contextuality as a valuable resource for state discrimination, we haven't take real environment into consideration. Thus, the next solid step to the final application of this resource could be a detailed study upon its performance in noisy channel\cite{rossi2023contextuality}, or PT-broken system\cite{chen2022quantum} as well as other non-Hermitian cases. Second, the method used here can also be applied to find and realize a even simpler version for testing contextuality, using the performance of QSD tasks as criteria, as suggested in \cite{zhang2022experimental}. Last but not least, our study contributes to not only the fundamental theory of quantum mechanics, but also practical quantum information tasks, by opening the door to demonstrate quantum advantage of various other tasks in which QSD tasks serve as a basic and key unit, such as quantum key distribution\cite{leifer2005pre,pusey2014anomalous}, randomness generation\cite{brask2017megahertz,i2022quantum}, and quantum machine learning\cite{lloyd2020quantum}. 
\section{ACKNOWLEDGEMENT}
This work was supported by the National Natural Science Foundation of China (No. 12004358); National Natural Science Foundation Regional Innovation and Development Joint Fund (No. U19A2075); Fundamental Research Funds for the Central Universities (No. 202041012, No. 202364008, No. 841912027); Natural Science Foundation of Shandong Province (No. ZR2021ZD19); Young Talents Project at Ocean University of China (No. 861901013107).
\appendix
\bibliography{universalsample}

\end{document}